# MULTIRESOLUTION FULLY CONVOLUTIONAL NETWORKS
# TO DETECT CLOUDS AND SNOW THROUGH OPTICAL SATELLITE IMAGES

Debvrat Varshney[1*], Claudio Persello[2], Prasun Kumar Gupta[3], Bhaskar Ramachandra Nikam[3]

[1]University of Maryland Baltimore County, United States of America
[2]Faculty of Geo-Information Science and Earth Observation (ITC), University of Twente, Netherlands
[3]Indian Institute of Remote Sensing (IIRS), Dehradun, India
*Corresponding author: dvarshney@umbc.edu, work performed as a Masters student during the IIRS-ITC Joint Education Program

***Abstract:*** *Clouds and snow have similar spectral features in the visible and near-infrared (VNIR) range and are thus difficult to distinguish from each other in high resolution VNIR images. We address this issue by introducing a shortwave-infrared (SWIR) band where clouds are highly reflective, and snow is absorptive. As SWIR is typically of a lower resolution compared to VNIR, this study proposes a multiresolution fully convolutional neural network (FCN) that can effectively detect clouds and snow in VNIR images. We fuse the multiresolution bands within a deep FCN and perform semantic segmentation at the higher, VNIR resolution. Such a fusion-based classifier, trained in an end-to-end manner, achieved 94.31% overall accuracy and an F1 score of 97.67% for clouds on Resourcesat-2 data captured over the state of Uttarakhand, India. These scores were found to be 30% higher than a Random Forest classifier, and 10% higher than a standalone single-resolution FCN. Apart from being useful for cloud detection purposes, the study also highlights the potential of convolutional neural networks for multi-sensor fusion problems.*

***Keywords:*** *Cloud Detection, Snow Cover Mapping, Deep Learning, Multiresolution Fusion, Fully Convolutional Networks*

## 1. INTRODUCTION

Snow is an important feature of our environment. It helps in balancing the heat flow between the Earth surface and atmosphere. Its presence in a basin also affects surface moisture, thereby contributing to water runoff [1]. Analyzing the snow cover area (SCA) plays an extensive role in managing hydrological resources for agricultural & societal needs [2], [3]. Moreover, snow cover maps are heavily utilized for climate studies, especially in quantifying glacier mass balance [4], [5]. Such spatial understanding of snow has historically been made through laborious physical surveys, which are mainly point measurements, and thus do not provide good estimates of the entire areal cover. Further, as snow is present on mountainous terrain, which is rough and undulating, measurement excursions can easily translate into becoming an expensive and hazardous activity for human life. Satellite remote sensing helps mitigate this, as the vast extent and high spatial resolution captured by satellite images assist in making accurate snow cover assessments.

The spatial extent of snow is mapped through optical images, as they provide an advantage in capturing snow reflectivity [6], [7]. However, optical satellite images are generally obstructed by cloud cover. Clouds, having similar reflectance as snow in the visible and near infrared (VNIR) range, become quite a hindrance when mapping snow in these spectral bandwidths [8]. In this regard, the shortwave infrared (SWIR) becomes a better alternative to discriminate clouds and snow [5]. The lower reflectance of snow, as compared to clouds, in the SWIR range can be used to improve SCA estimates (Figure 1).

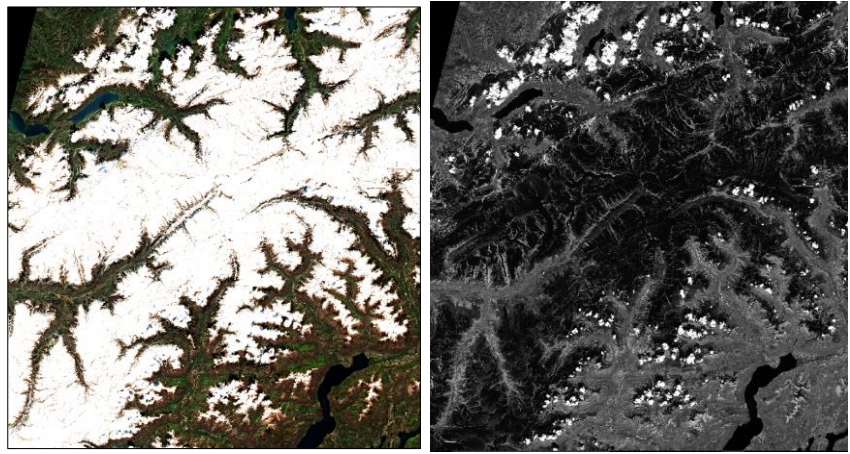

Figure 1: A true color image (left) of the Swiss Alps, acquired by Sentinel-2A. Such an image is ineffective for snow assessment as it has cloud cover. The SWIR channel of the same satellite (right) highlights clouds with bright pixels.

As infrared images typically have low resolution and poor texture [9], our study focuses on building a fusion-based algorithm combining SWIR-VNIR bands for snow and cloud classification. Convolutional Neural Networks are explored due to their recent success in image recognition challenges [10], [11] and specifically optical remote sensing [12], [13]. We employ the properties of such networks to fuse data from the multiresolution Linear Imaging Self Scanners (LISS) on-board Resourcesat-2 satellite (RS2), extract features from it, and perform pixelwise classification to obtain a high-resolution cloud mask over snow covered regions. The entire procedure is performed in a single, end-to-end framework and is void of any pre- or post-processing steps. Such an automated algorithm can be used to build native cloud masks for Earth Observation datasets from Indian Remote Sensing (IRS) satellites, which has been lacking until now.

## 2. BACKGROUND

### 2.1. Convolutional Neural Networks

Convolutional Neural Networks (ConvNet) are a special type of neural networks, where the neurons perform a convolution. Each neuron of a ConvNet is at a time exposed to only a small region of the input data and takes its weighted sum. It then convolves, or slides onto a neighboring region with the same set of weights, eventually covering the entire input data layer. This procedure is unlike regular neural networks, such as the multi-layer perceptron (MLP), where every hidden neuron has weighted connections to the *entire* input data layer. Hence, the number of learnable weights required in a ConvNet are far less compared to an MLP. The utility of this property becomes more apparent for high dimensional datasets.

The activations applied over the weighted sums are non-linear functions like sigmoid and hyperbolic tangents; so that the output is differentiable and the weights can be learned through the gradient descent algorithm [14]. Recently, Rectified Linear Units (ReLU, Equation 1a) have been popularly used as activations due to their advantage in the vanishing gradient problem and faster training time. Modified versions of these activations, such as leaky ReLU (Equation 1b) were used in this study due to their superior performance [15]. In these equations, $x$ denotes the weighted sum, $y$ its corresponding activation and $\alpha$ is a constant parameter.

$$y = f(x) = \begin{cases} x, & x \geq 0 \\ 0, & x < 0 \end{cases} \tag{1a}$$

$$y = f(x) = \begin{cases} x, & x \geq 0 \\ \alpha x, & x < 0, \alpha \geq 0 \end{cases} \tag{1b}$$

Such ConvNets with multiple hidden layers provide hierarchical feature learning. As the convolving neurons (or filters) only focus on a small area at a time, a convolutional layer helps in understanding the local relationship between pixels. The initial layers of the network extract basic features such as edges first; and patterns from these layers are later pooled together to form more complex objects, among the deeper layers. Comparatively, all the fully connected layers of an MLP look at the previous layer in its entirety, trying to understand a more global relationship among pixels. Thus, multiple layers of a deep ConvNet can detect more intricate patterns and features from the data, as compared to an MLP. This feature extraction capability of ConvNets make them useful for computer vision tasks such as object detection and image classification.

To classify each pixel of the image as cloud or snow, we use Fully Convolutional Networks (FCN). FCNs do not have fully connected layers in the end, and are trained in an end-to-end manner on pixel-wise labels of the input image. In this way, rather than producing a single label for an entire image, they produce a label for every pixel of the image. Thus they assist in semantic segmentation [16]. In our case, we train such networks with spatio-contextual information from a variety of snow cover images, and use them to classify unseen areas. Further, in order to use the multiresolution SWIR and VNIR bands, we resample them using a series of convolution operations to fuse (concatenate) them within the FCN architecture. Such an FCN-based-fusion approach has shown a nearly 5% increase in classification accuracies compared to standard fusion techniques being followed by an FCN classification [17].

### 2.2. Cloud Detection Algorithms

A current state-of-the-art algorithm for cloud detection is Fmask developed in [18]. The algorithm was built to detect thick and semi-transparent clouds, along with their shadows on Landsat imagery. This algorithm was found to be very effective; and as the Landsat images are freely available and popularly used, the algorithm soon became an asset to the remote sensing community. The usefulness of this algorithm further led to its improvement and application on Sentinel-2 images [19]. The improved algorithm detects clouds by a series of spectral tests to generate a cloud probability mask, while it uses thresholds on Normalized Difference Snow Index (NDSI) and Brightness Temperature to create a snow layer. The algorithm further detects cloud shadows by multiple geometrical techniques, which rely on satellite's metadata such as the sensor's view angle, solar zenith angle etc. Altogether, Fmask applies a scene based threshold to all the pixels in a neighborhood, and classifies the pixels into clouds, cloud shadows, and snow; in that priority. It fails to understand the spectral-spatial difference among the varying class objects on its own, as it majorly depends on the threshold values which have been applied.

As such an algorithm is sensor specific, needs associated metadata, and majorly uses spectral thresholds for cloud determination; there lies a scope to build algorithms which are easily adaptable across multiple sensors and robust enough to detect clouds and snow in a spatio-contextual sense. Convolutional Neural Networks, help in such a feature detection scenario, as described in the previous section. Their feature detection capability has been employed in several works to create effective cloud masks [20]–[23].

The authors in [20] have used exhaustive amounts of data to train deep neural nets for clouds, cloud shadows, and snow detection. But these networks are not convolutional and require post-processing for snow-cloud correction. Similarly, the authors in [21] have incorporated deep neural nets for cloud detection, but they have not developed them specifically for snow-covered areas as they fail to distinguish between clouds and snow accurately. The authors of [22] have developed a ConvNet for detecting clouds, but then again use a separate snow/ice removal framework in the pre-processing stage. Furthermore, authors of [23] have developed a deep convolutional network for distinguishing clouds from snow, which majorly uses a multiscale prediction strategy, combining low-level feature maps with high-level feature maps. The intermediate feature maps are of varying spatial resolutions in this case, and they interpolate these maps separately before combining them. In this paper, we propose an approach based on a deep ConvNet that takes as input both high resolution VNIR images and a corresponding lower resolution SWIR band. This work is an extension of [24] and contains more quantitative analysis, apart from comparing with an additional classifier. By resampling and fusing the SWIR band with VNIR images within the FCN, our ConvNet improves the detection of clouds over snow covered regions, without requiring a separate snow/ice removal step. The main contribution of our proposed ConvNet architecture is that it performs semantic segmentation for snow and clouds, using multiresolution data, all within a single, end-to-end classifier.

## 3. METHODOLOGY

As our approach involves extracting features from bands of varying resolutions, this can be performed by building an FCN over a product fused through traditional techniques of pansharpening or interpolation. This process, involving separate stages of fusion and feature extraction, was recently compared against a single-stage framework called FuseNet [17]. The performance of this multiresolution ConvNet was found to surpass traditional techniques, and we build a similar multi-armed architecture to concatenate low-resolution SWIR features with the downsampled and convoluted VNIR features. Subsequent sections describe our network architecture in detail.

### 3.1. Network Characteristics

We use convolutional layers of dimensions $F \times F \times C \times K$, where $K$ is the number of filters, each having $C$ channels of size $F \times F$. Each filter convolves with a stride $S$, both in the horizontal and vertical directions, to produce an output feature map of lower dimension. To create feature maps of the same dimension, the input feature map is generally padded with additional zero-valued rows and columns on the outskirts of the image matrix. We use the same number of rows as columns for padding, and denote the number of such rows with $P$. Hence, for any input image patch having $C$ channels of size $M \times M$, our convolutional layer will produce $K$ output feature maps of size $M' \times M'$, where $M'$ is given as:

$$M' = \frac{M - F + 2P}{S} + 1 \qquad (2)$$

We also use max pooling layers to resample the VNIR bands to the SWIR's resolution, before concatenation. Max pooling gives the maximum activations from a given filter and using such non-padded filters with a stride equal to their filter width, results in downsampling the input feature map. Lastly, we use transposed convolutions on the fused bottleneck. They function in a reverse manner compared to regular convolutions by making the input matrix sparse i.e. they add multiple zeros in between the matrix elements. Hence, they upsample the input matrix while extracting features at the same time. Such a layer creates an output feature map of dimensions $M'\times M'$, given by Equation 3, where $p$ is the cropping factor, and all other terms have the same meaning as used earlier. We use these layers to upsample the fused bottleneck and extract features from it.

$$M' = S(M - 1) + F - 2p \qquad (3)$$

For all the convolutional layers, we use $S=1$ and appropriate padding to keep the dimensions of input and output feature maps the same. For some convolutional layers, we also apply a dilation factor to the filter. The dilation is achieved by inserting zeros in between the weights of the filter, so as to increase its size by a factor $d$ as given in Equation 4. Doing so increases the spatial support of the filter, without increasing the number of trainable weights per layer [25].

$$F' = d(F - 1) + 1 \qquad (4)$$

### 3.2. Learning Algorithm

We train our ConvNet by minimizing a cross entropy loss function as given in Equation 5. The objective of the training is to reduce the error computed by the loss function, for a weight vector $\boldsymbol{w}$ used by the network.

$$E_N(\boldsymbol{w}) = -\sum_i^N \boldsymbol{z}_i \cdot log(\boldsymbol{y}_i) \qquad (5)$$

Here, $N$ is the number of reference training samples (pixels) in a mini-batch, $\boldsymbol{z}$ and $\boldsymbol{y}$ are vectors having size equal to the number of classes in the input image. $\boldsymbol{z}$ is made up of zeroes, except at the index corresponding to the pixel's labeled class, which has a value of one. $\boldsymbol{y}$ is made up of normalized values coming from softmax activation, used as the final classification layer and given in Equation 6. Here $x_m$ represents the $m$th channel output of the last convolutional layer, and $y_m$ represents its associated activation. The last convolutional layer has a total of $k$ channels, representing the number of classes.

$$y_m = f(x_m) = \frac{e^{x_m}}{\sum_{j=1}^k e^{x_j}} \qquad (6)$$

The training takes place through a mini-batch gradient descent, where weights get modified after each epoch, in the direction of decreasing error. The updated weight at epoch $t$ is given in Equation 7, where $\eta$ represents the learning rate and $\alpha$ is the momentum, both varying between 0 and 1. Such gradient descent methods have shown better generalization ability than adaptive methods [26].

$$\Delta \boldsymbol{w}_t = -\eta \frac{\partial E_N(\boldsymbol{w})}{\partial \boldsymbol{w}_t} + \alpha \Delta \boldsymbol{w}_{t-1} \qquad (7)$$

To avoid overfitting, the loss function of Equation 5 is penalized by a squared L$^2$ norm of the weight vector $\boldsymbol{w}$. The contribution of this norm is controlled by a parameter $\lambda$, known as weight decay. The modified loss function is given as $Q_N(\boldsymbol{w})$ in Equation 8.

$$Q_N(\boldsymbol{w}) = E_N(\boldsymbol{w}) + \lambda \|\boldsymbol{w}\|_2^2 \qquad (8)$$

### 3.3. Proposed Architecture and its Comparisons

We built our ConvNet using the network parameters described in Section 3.1. Every convolutional and transposed convolutional layer of our network was followed by a Batch Normalization (BN) layer. The BN layers after every convolutional layer are further followed by a leaky ReLU, with $\alpha$=0.1 (Equation 1b). We downsample the VNIR bands twice (because of our dataset; see Section 4.1) through max-pool layers and concatenate them with the lower resolution SWIR band. As we want the final predictions to be at the higher (VNIR) resolution, we employ transposed convolutions to upsample the fused composite. We refer to this cloud-and-snow detecting network as CloudSNet, shown in Figure 2.

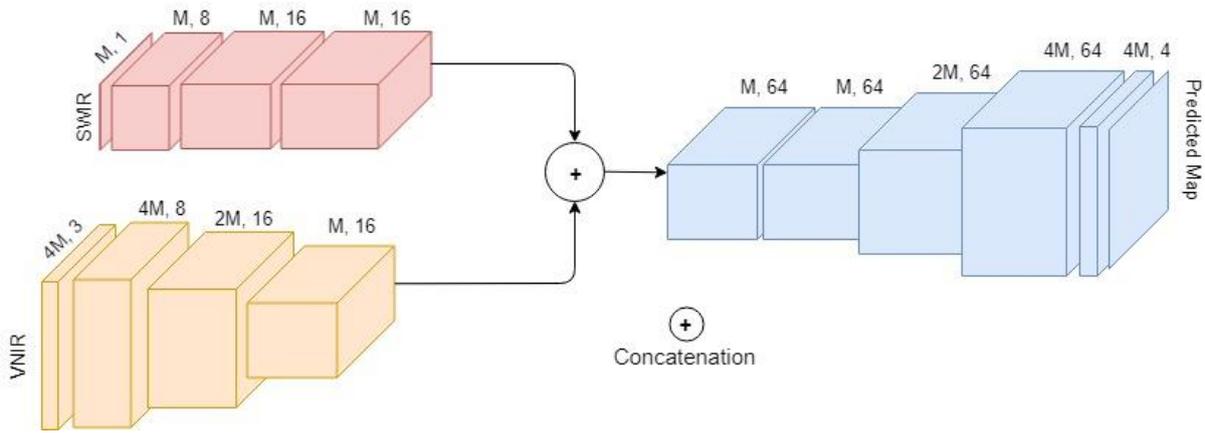

Figure 2: CloudSNet - for cloud and snow discrimination. VNIR bands are subsampled and fused with SWIR, with the final composite getting upsampled to the higher resolution for pixel-wise classification. Here, M denotes a square image patch of size M×M. Moreover, features indicated as, say, (4M, 8) imply eight channels of dimension 4M×4M.

To analyze the utility of SWIR fusion for cloud detection, we built two more FCNs having similar parameters as CloudSNet but working on single resolution data. That is, one worked only on VNIR bands, and the other only the SWIR band. We refer to these networks as FCN$_{VNIR}$ (Figure 3) and FCN$_{SWIR}$ (Figure 4), respectively. Table 1 gives all three network architectures in detail. We further compare our ConvNets with a Random Forest (RF) classifier, which is a standard, pixel-based machine learning algorithm. This classifier was trained only on the VNIR bands, and the comparison helps us analyze the utility of spatio-contextual information for cloud detection purposes, especially over snow covered regions.

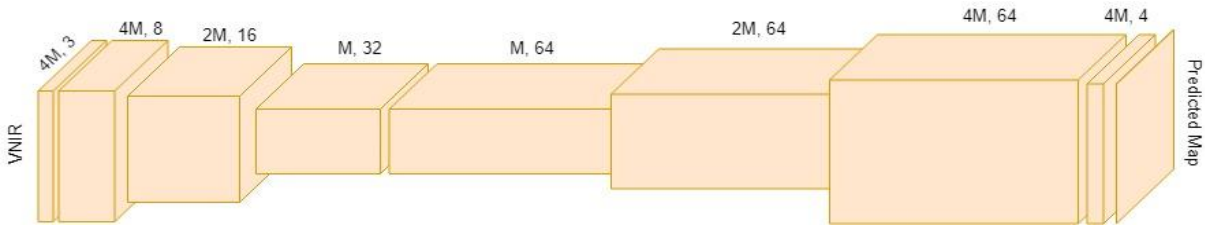

Figure 3: FCN$_{VNIR}$ working solely on the VNIR bands. Network parameters are similar to those being used by CloudSNet on its VNIR bands

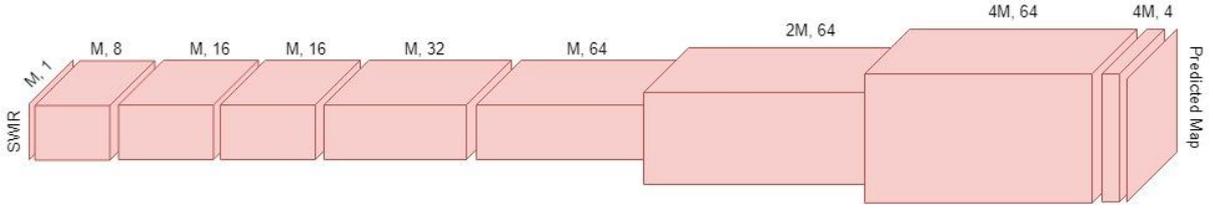

Figure 4: FCN<sub>SWIR</sub> working solely on a SWIR band. Network parameters are similar to those used by CloudSNet on its SWIR band.

Table 1: Architectural details of CloudSNet, FCN$_{VNIR}$, and FCN$_{SWIR}$

| CloudSNet | | FCN$_{VNIR}$ | FCN$_{SWIR}$ |
|---|---|---|---|
| **VNIR** | **SWIR** | | |
| Conv5-1-8 maxpool | Conv1-1-8 | Conv5-1-8 maxpool | Conv1-1-8 |
| Conv5-1-16 maxpool | Conv3-1-16 | Conv5-1-16 maxpool | Conv3-1-16 |
| | Conv5-1-32 | Conv5-1-32 | Conv5-1-16 |
| **Concatenation** | | Conv5-2-64 | Conv5-1-32 |
| Conv5-1-64 | | TConv4-2-1-64 | Conv5-2-64 |
| Conv5-2-64 | | TConv4-2-1-64 | TConv4-2-1-64 |
| TConv4-2-1-64 | | Conv1-1-4 | TConv4-2-1-64 |
| TConv4-2-1-64 | | | Conv1-1-4 |
| Conv1-1-4 | | | |

In Table 1, every convolutional layer is represented as Conv<filter width>-<dilation factor>-<number of filters>. Example: a Conv5-2-64 layer means 64 filters of size 5×5, with a dilation factor of 2. All convolutional filters move with a stride of 1. Appropriate padding was applied to keep the dimensions of the input and output feature maps the same. Max pooling layers are represented as maxpool, all of them having a 2×2 window, moving with a stride of 2. Transposed Convolutions are represented as TConv<filter width>-<stride or upsampling factor>-<cropping factor>-<number of filters>. The filter width, the upsampling factor, and the cropping factor are related as in Equation 3. We wanted to segment the image into four classes: clouds, snow, shadows, and rest; and hence kept four output channels in the final convolutional layer.

## 4. DATA AND EXPERIMENTAL SETUP

### 4.1. Sensors and Data

We carried out our experiments using LISS-III and IV sensors present onboard RS2. This satellite is majorly meant for monitoring crops, providing assistance to farmers, managing water resources, having all its sensors looking at nadir. LISS-IV is a high resolution sensor (5m) having three bands (B2, B3, and B4) working in the VNIR range, whereas LISS-III is a medium resolution sensor (24m) consisting of a SWIR band (B5) along with VNIR bands. As ConvNets can only have integral strides, we cannot use these bands directly to bring them to a common resolution. Hence, we resampled B5 to 20m through Nearest Neighbour Interpolation before using it for our ConvNets. Note that this step is required only for our specific dataset, and is not required for the methodology in general. Further, we performed our tests on Path 97 Row 49 of RS2 (Figure 5), which covers a portion of the Indian state of Uttarakhand having some of the most vulnerable glaciers in the country. Our work can contribute here by making mapping and monitoring of glaciers more accurate.

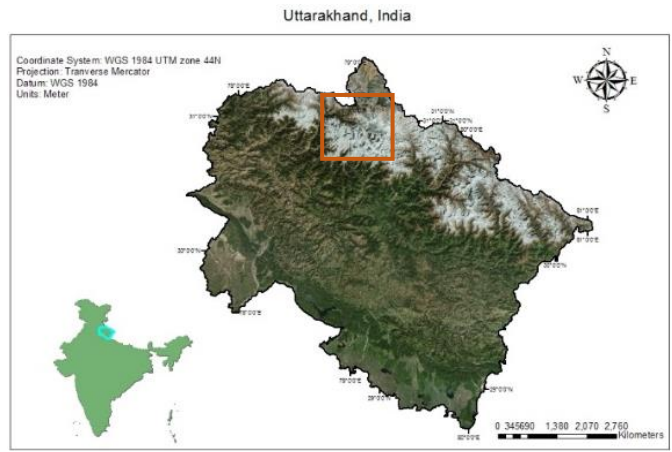

Figure 5: State of Uttarakhand in India. Area covered by Path 97, Row 49 of RS2 is highlighted in orange

We took two scenes from the specified path of RS-2 belonging to 9th October 2014 and 13th May 2015. Across these two scenes, we selected eight square tiles, of 100km² each. Four of these tiles were used for training our classifiers while four were used for testing purposes. The tiles had a dimension of 2000×2000 pixels on the false color composite of VNIR bands, and 500×500 pixels on the SWIR band. Figure 6 depicts the location of these tiles.

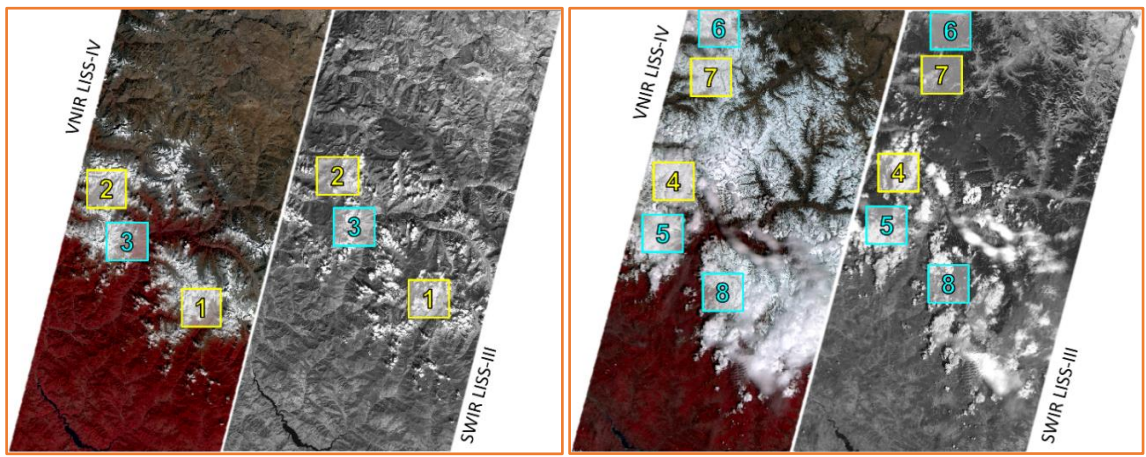

Figure 6: Study area on 9th Oct '14 (left) and 13th May '15 (right). Yellow marks the training tiles, while blue marks the test tiles across the multi-resolution data

### 4.2. Classifier Setup

We created reference labels at the higher (VNIR) resolution, for our supervised learning algorithms. This was done by visual interpretation on all the eight tiles. Apart from Clouds and Snow, we wanted to segment the data into two more classes – Shadows and Rest. Table 2 gives details of the reference pixel labels created.

Table 2: Number of reference pixels for every class, total number of reference pixels, and their usage purpose

| Clouds | Snow | Shadows | Rest | Total | Purpose |
|--------|------|---------|------|-------|---------|
| 2130885 | 1294481 | 326455 | 1023283 | 4775104 | Training + Validation |
| 2739517 | 1216855 | 297244 | 409306 | 4662922 | Test |

To train the ConvNets, we normalized the available dataset and extracted 2000 random patches from our training tiles. We also extracted 500 random patches from these tiles for validation purposes. We kept the size of each patch as 50×50 pixels (i.e. M=50) on the resampled SWIR band, and as this band was one-fourth the resolution of VNIR bands, kept the patch size as 200×200 pixels on the VNIR bands. We took the same sized patches i.e. 200×200 pixels from the reference tiles as well. We trained each of the ConvNets for 50 epochs, with a batch size of 32 patches. We reduced the learning rate logarithmically between $10^{-6}$ and $10^{-7}$, in steps equal to the number of epochs and analyzed the convergence of the loss function on the validation set to assess the ConvNets' performance. Further, we kept the weight decay as $5×10^{-4}$, momentum as 0.9 and performed Xavier initialization [27] for the filter weights.

To prepare the RF classifier on the VNIR bands, all the training labels were divided into eight batches, and every batch was trained by a separate forest made up of 400 trees, having information gain as the node splitting criterion. All the forests were combined to form the classifier.

### 4.3. Performance Metrics

We used the following metrics to analyze the performance of our ConvNets and RF classifier. These metrics were computed on the confusion matrix generated for the test set (cumulative of Tiles 3, 5, 6, and 8). The confusion matrix $C$, prepared for a total of $n$ = 4 classes, has columns corresponding to the reference pixels and rows corresponding to the predicted pixels. A matrix element denoted by $C_{ij}$ corresponds to the $j^{th}$ column of the $i^{th}$ row.

Overall Accuracy (OA): This is the total number of correctly predicted pixels, divided by the total number of labelled reference pixels. It is computed across all four classes of the confusion matrix and given by Equation 9 below.

$$OA = \frac{\sum_{i=1}^{n} C_{ii}}{\sum_{j=1}^{n} \sum_{i=1}^{n} C_{ij}} \tag{9}$$

F1 - Score: OA can be highly biased if there is an uneven class distribution in the image. Precision and recall, expressed in Equations 10 and 11 respectively for a class with label index $a$, help in this regard by highlighting the usefulness of a class' prediction.

$$Precision_a = \frac{C_{aa}}{\sum_{j=1}^{n} C_{aj}} \tag{10}$$

$$Recall_a = \frac{C_{aa}}{\sum_{i=1}^{n} C_{ia}} \tag{11}$$

Precision is the number of correctly predicted pixels of a class, divided by the total number of pixels predicted as that class. On the other hand, recall is the number of correctly predicted pixels of a class, divided by the total number of reference pixels for that class. We use the harmonic mean of precision and recall, known as the F1 score, to assess our classifiers' performance. The formula for F1 score is expressed in Equation 12, and we compute this metric only for the snow and cloud classes.

$$F1_a = \frac{2 \cdot Precision_a \cdot Recall_a}{Precision_a + Recall_a} \tag{12}$$

<u>Average F1 Score</u>: To assess the performance of detecting snow and clouds collectively, we first calculate the precision and recall micro-averaged over these two classes, and given in Equations 13 and 14, respectively. Here, $s$ defines the index label for snow and $c$ defines the index label for clouds. We term the harmonic mean of such micro averages as the Average F1 score.

$$Precision_\mu = \frac{C_{ss} + C_{cc}}{\sum_{j=1}^{n} C_{sj} + \sum_{j=1}^{n} C_{cj}} \tag{13}$$

$$Recall_\mu = \frac{C_{ss} + C_{cc}}{\sum_{i=1}^{n} C_{is} + \sum_{i=1}^{n} C_{ic}} \tag{14}$$

## 5. RESULTS AND DISCUSSION

This section discusses the quantitative and qualitative performance of our classifiers. From Table 3, we see that CloudSNet gives an advantage for cloud detection (over snow) compared to regular FCNs, while RF underperforms across all the metrics. We also see that SWIR-based FCNs give more than 12% higher recall for clouds and 15% higher precision for snow, as compared to FCN$_{VNIR}$; while all three ConvNets give nearly the same precision for clouds. Moreover, FCN$_{SWIR}$ gives a 4% smaller recall for snow compared to the other two networks, which perform almost the same in this regard. Altogether, SWIR based FCNs give higher F1 scores compared to an FCN on VNIR bands alone, with CloudSNet outperforming the other classifiers by at least 2% across most of the metrics. Moreover, CloudSNet improves upon the Overall Accuracy, micro averaged Precision and micro averaged Recall of FCN$_{VNIR}$ by nearly 10%.

Table 3: Major performance metrics (in %) of the classifiers

| Classifier | Clouds | | | Snow | | | Clouds-and-Snow | | | |
|---|---|---|---|---|---|---|---|---|---|---|
| | Precision | Recall | F1 | Precision | Recall | F1 | Precision$_\mu$ | Recall$_\mu$ | Avg F1 | OA |
| RF | 77.85 | 56.54 | 65.51 | 40.73 | 63.31 | 49.57 | 59.76 | 58.62 | 59.19 | 61.02 |
| FCN$_{SWIR}$ | 98.06 | 94.80 | 96.40 | 87.57 | 86.90 | 87.23 | 94.78 | 92.37 | 93.56 | 88.88 |
| FCN$_{VNIR}$ | 96.28 | 82.85 | 89.07 | 72.16 | **90.94** | 80.47 | 86.77 | 85.34 | 86.05 | 85.99 |
| CloudSNet | **98.45** | **96.90** | **97.67** | **94.73** | 90.40 | **92.51** | **97.33** | **94.90** | **96.10** | **94.31** |

Table 4: Minor performance metrics (in %) of the classifiers

| Classifier | Shadows | | | Rest | | |
|---|---|---|---|---|---|---|
| | Precision | Recall | F1 | Precision | Recall | F1 |
| RF | 76.86 | 42.87 | 55.04 | 64.71 | 97.36 | 77.74 |
| FCN$_{SWIR}$ | 55.82 | 48.35 | 51.82 | 62.97 | 84.53 | 72.17 |
| FCN$_{VNIR}$ | 80.72 | 78.05 | 79.36 | **82.81** | **98.00** | **89.77** |
| CloudSNet | **80.89** | **86.25** | **83.49** | 79.21 | 94.50 | 86.18 |

From the confusion matrix of FCN$_{VNIR}$ (Table 5), we see that 15.25% of actual cloud pixels are predicted as snow, while 1.90% are predicted as shadows or rest. These numbers decrease to 1.64% and 1.46% respectively in CloudSNet (Table 6). Out of the labelled snow pixels, 3.46% were misclassified as clouds by FCN$_{VNIR}$, while this ratio reduced to 0.48% in case of CloudSNet. Both these observations highlight the advantage that SWIR based resampling brings in detecting clouds, and reducing their misclassification as snow.

Table 5: Confusion Matrix of FCN$_{VNIR}$. Each column shows the percentage of labelled pixels predicted as a specific class. 'Total' values are the number of pixels.

| Class | Clouds | Snow | Shadows | Rest | Total |
|---|---|---|---|---|---|
| Clouds | **82.85** | 3.46 | 13.50 | 1.32 | 2357500 |
| Snow | 15.25 | **90.94** | 2.75 | 0.26 | 1533598 |
| Shadows | 0.65 | 2.95 | **78.05** | 0.42 | 287386 |
| Rest | 1.25 | 2.65 | 5.70 | **98.00** | 484438 |
| Total | 2739517 | 1216855 | 297244 | 409306 | 4662922 |

Table 6: Confusion Matrix of CloudSNet. Each column shows the percentage of labelled pixels predicted as a specific class. 'Total' values are the number of pixels.

| Class | Clouds | Snow | Shadows | Rest | Total |
|---|---|---|---|---|---|
| Clouds | **96.90** | 0.48 | 4.61 | 5.46 | 2696381 |
| Snow | 1.64 | **90.40** | 5.44 | 0.01 | 1161325 |
| Shadows | 0.51 | 3.82 | **86.25** | 0.03 | 316926 |
| Rest | 0.95 | 5.30 | 3.70 | **94.50** | 488290 |
| Total | 2739517 | 1216855 | 297244 | 409306 | 4662922 |

Coming to FCN$_{SWIR}$ in Table 7, we see that nearly 4.7% of the actual cloud pixels are misclassified into the Rest class, while this happens for only 0.95% of the clouds in case of CloudSNet. Moreover, FCN$_{SWIR}$ misclassifies nearly 9% of the snow as shadows, while CloudSNet does so for 3.82% of the snow. The major reason for these misclassifications is that in SWIR, the 'Rest' regions, covering features such as moraines and vegetation, are also bright and similar to clouds. Also, snow appears dark in this spectral range and is easily confused as shadows. Having VNIR bands becomes handy in such a case as these are spectrally more diverse. This is seen in the higher F1 scores achieved by VNIR-based classifiers for Shadows and Rest, as compared to a SWIR-only classifier (Table 4). Hence, even though FCN$_{SWIR}$ involves lesser computations and is reasonably accurate in detecting snow and clouds, it does not become as optimum a choice as CloudSNet.

Table 7: Confusion Matrix of FCN$_{SWIR}$. Each column shows the percentage of labelled pixels predicted as a specific class. 'Total' values are the number of pixels.

| Class | Clouds | Snow | Shadows | Rest | Total |
|---|---|---|---|---|---|
| Clouds | **94.80** | 0.03 | 1.11 | 11.64 | 2648477 |
| Snow | 0.42 | **86.90** | 41.83 | 3.49 | 1207547 |
| Shadows | 0.11 | 8.99 | **48.35** | 0.34 | 257446 |
| Rest | 4.67 | 4.08 | 8.71 | **84.53** | 549452 |
| Total | 2739517 | 1216855 | 297244 | 409306 | 4662922 |

Table 8: Confusion Matrix of RF. Each column shows the percentage of labelled pixels predicted as a specific class. 'Total' values are the number of pixels.

| Class | Clouds | Snow | Shadows | Rest | Total |
|---|---|---|---|---|---|
| Clouds | **56.54** | 32.43 | 14.00 | 1.10 | 1989786 |
| Snow | 39.58 | **63.31** | 11.97 | 0.31 | 1891462 |
| Shadows | 0.38 | 1.88 | **42.87** | 1.23 | 165788 |
| Rest | 3.50 | 2.38 | 31.16 | **97.36** | 615886 |
| Total | 2739517 | 1216855 | 297244 | 409306 | 4662922 |

Table 8 further shows the confusion matrix for the Random Forest classifier for reference. This classifier underperforms across all the metrics (Table 3), highlighting the inability of a pixel-based machine learning classifier for our problem statement. Moreover, the output given by this classifier is full of salt-and-pepper-noise, overestimating snow in the images (Figure 7). Hence, Random Forest does not become practically useful for our purpose.

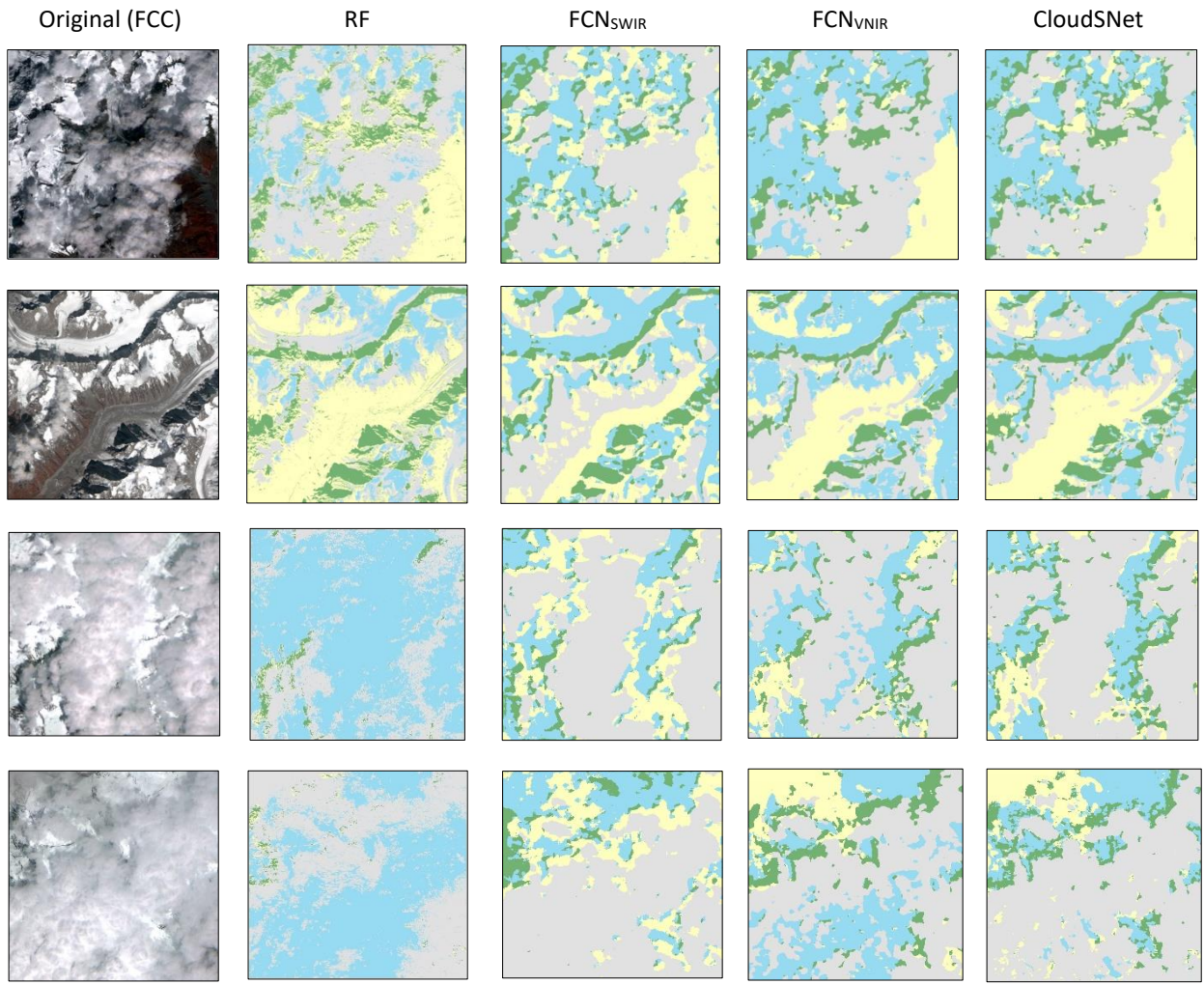

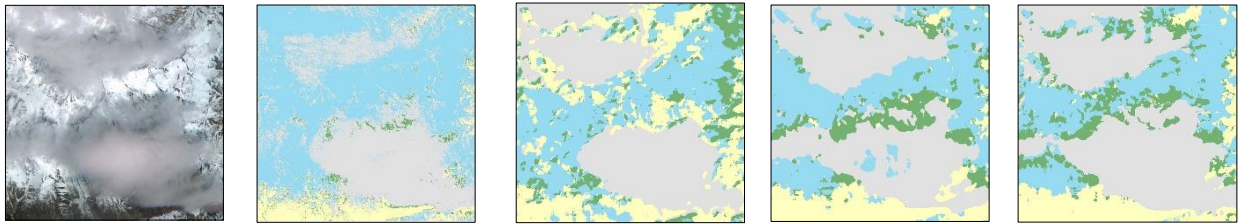

Figure 7: False colour composite (FCC) of the original LISS-IV image, outputs from RF, FCN<sub>SWIR</sub>, FCN<sub>VNIR</sub> and CloudSNet for Tiles 1, 2, 4, 5 and 6. Legend: Grey = Clouds , Blue = Snow , Green = Shadows , Yellow = Rest of the region.

## 6. CONCLUSION AND FUTURE WORK

In this work, we presented a fully convolutional multi-resolution network for distinguishing clouds from snow. This was achieved by downsampling VNIR bands through max pooling, fusing them with lower resolution SWIR band, and upsampling the composite back to the higher resolution through a series of transposed convolutional layers. Such a SWIR-and-VNIR based FCN achieved higher F1 scores for clouds and snow, as compared to standalone single-resolution FCNs. Moreover, the higher performance of FCNs compared to a standard pixel-based machine learning classifier (Random Forest), highlights the advantage that spatio-contextual learning brings for cloud detection problems and semantic segmentation. Such deep learning classifiers become more promising, especially over large datasets, as compared to traditional pixel-based cloud detection algorithms which require feature engineering. Moreover, our algorithm is completely independent of the sensor's metadata and only requires reflectance values as input.

Although our specific dataset involved a pre-processing interpolation stage, such a step would not be required for other datasets such as Sentinel-2 MSI, where the SWIR and VNIR resolutions are integral multiples of each other. Moreover, through the proposed method, we can fuse cirrus bands to separate out thin clouds, and thermal infrared bands for increasing the accuracy of snow edge detection. As our work involved manual labelling of large datasets, one can instead use Generative Adversarial Networks for performing unsupervised classification. Such methods, requiring high processing power, can have VNIR bands as the Generator and SWIR as the Discriminator Network for clouds and snow classification. Moreover, SAR data can be incorporated to especially detect snow which is present beneath the cloud cover, and thus assist in creating accurate cloud-free snow cover maps.

## ACKNOWLEDGEMENTS

The authors would like to thank National Remote Sensing Centre, under the Department of Space, Government of India, for providing the data captured by Resourcesat-2. Processing and analysis of data was carried out at the Indian Institute of Remote Sensing, Dehradun.

## COMPETING INTERESTS

The authors declare no competing interests